\documentclass[aps,twocolumn,nofootinbib]{revtex4}

\usepackage{graphicx}
\usepackage{dcolumn}
\usepackage{bm}

\usepackage{amssymb,amsfonts}
\usepackage{epsfig}
\usepackage{epstopdf}
\usepackage[all]{xy}
\usepackage{amsthm}
\usepackage{dcolumn}
\usepackage{hyperref}
\usepackage{url}

\newcommand{\be}{\begin{equation}}
\newcommand{\ee}{\end{equation}}
\newcommand{\bea}{\begin{eqnarray}}
\newcommand{\nn}{\nonumber}
\newcommand{\eea}{\end{eqnarray}}

\def\inbar{\,\vrule height1.5ex width.4pt depth0pt}
\def\IR{\relax{\rm I\kern-.18em R}}
\def\IC{\relax\hbox{$\inbar\kern-.3em{\rm C}$}}

\begin{document}

\title{Examining a covariant and renormalizable quantum field theory in de Sitter space\\ by studying "black hole radiation"}

\author{Hamed Pejhan\footnote{h.pejhan@piau.ac.ir}}
\affiliation{Department of Physics, Science and Research Branch, Islamic Azad University, Tehran, Iran}
\author{Surena Rahbardehghan\footnote{s.rahbardehghan@iauctb.ac.ir}}
\affiliation{Department of Physics, Islamic Azad University, Central Branch, Tehran, Iran}

\begin{abstract}
Respecting that any consistent quantum field theory in curved space-time must include black hole radiation, in this paper, we examine the Krein-Gupta-Bleuler (KGB) formalism as an inevitable quantization scheme in order to follow the guideline of the covariance of minimally coupled massless scalar field and linear gravity on de Sitter (dS) background in the sense of Wightman-G\"{a}rding approach, by investigating thermodynamical aspects of black holes. The formalism is interestingly free of pathological large distance behavior. In this construction, also, no infinite term appears in the calculation of expectation values of the energy-momentum tensor (we have an automatic and covariant renormalization) which results in the vacuum energy of the free field vanishes. However, the existence of an effective potential barrier, intrinsically created by black holes gravitational field, gives a Casimir-type contribution to the vacuum expectation value of the energy-momentum tensor. On this basis, by evaluating the Casimir energy-momentum tensor for a conformally coupled massless scalar field in the vicinity of a non-rotating black hole event horizon through the KGB quantization, in this work, we explicitly prove that the hole produces black-body radiation which its temperature exactly coincides with the result obtained by Hawking for black hole radiation.
\end{abstract}
\maketitle

\section{Introduction}
\label{sec:intro}
Black holes, on the one hand, are indeed the simplest predictions of the general theory of relativity, and on the other hand, they are the paradigmatic objects to test possible quantum theories of gravity. More accurately, respecting the celebrated Hawking prediction of black hole evaporation, they possess thermodynamical properties \cite{Hawking199}, whose statistical origin should be characterized by a quantum theory of gravity. In the absence of such a complete theory (quantum gravity), however, quantum field theory (QFT) in curved space-time is of great importance and has been extensively investigated in the literature (\emph{e.g.} see \cite{Birrell1982,Wald1984}). A fundamental aspect of any consistent QFT in curved space is therefore the approach of the theory to black hole radiation. Indeed, how one can study black hole thermodynamics in the context of a theory and regarding its specifications must be answered. De Sitter space-time, a maximally symmetric solution of the vacuum Einstein equations with positive cosmological constant, is one of the most popular curved space-time, especially because of its relevance to the inflationary cosmology \cite{Kazanas59,Sato195,Guth347,Linde389,Albrecht1220}. Moreover, the recent astronomical data which are strongly in favor of an accelerating universe, indicate that the universe has a positive cosmological constant and may approach to de Sitter phase asymptotically \cite{riess}. It is also attracting to peruse physics in de Sitter space due to the dS/CFT correspondence \cite{Strominger034}. Note that, even on this simple space-time, we encounter with difficulties in quantizing fields.

As is well-known, the minimal requirements for an acceptable quantization procedure of a field are: the field must satisfy the field equation, commute for space-like separations (micro-causality) and verify the covariance condition (the Fock space being closed under the action of the isometry group). Respecting these physical requirements, however, the usual canonical quantization in which the Fock space is constructed over the Hilbert space (the scalar product is positive) fails in many situations. A famous example is the so-called de Sitter minimally coupled massless scalar field, for which, Allen has proved that a Hilbertian dS-invariant Fock vacuum does not exist. Therefore, the covariant canonical quantization cannot be constructed over the Hilbert space \cite{Allen3136,Allen3771}. The origin of the problem is indeed in the appearance of infrared divergence, which in turn is due to the existence of a constant solution of the field. Allen has expressed that this constant function can be regarded as the real part of a zero mode. Although the norm of this zero mode is positive, inclusion in the Hilbertian structure of the one-particle sector is not possible. More precisely, the de Sitter group action on this mode creates all the negative frequency solutions (with respect to the conformal time) of the field equation. Therefore, the constructed Hilbertian Fock space over any complete set of modes including the zero mode (${\cal{H}}_{+}=\{\sum_{k\geq0} \alpha_k\phi_k; \sum_{k\geq0}|\alpha_k|^2<\infty \}$, $\phi_k$ is defined in \cite{Gazeau1415}) is not closed under the dS group action. Indeed, the crucial point about the minimally coupled field is originated in the fact that a covariant decomposition $ {\cal{H}}_{+} \oplus {\cal{H}}_{-}$ does not exist; none of ${\cal{H}}_{+}$ and ${\cal{H}}_{-}$ (anti-Hilbert space with a negative definite inner product) carry the unitary irreducible representations of the de Sitter group.\footnote{In the case of the scalar massive field, however, such a decomposition exists, where ${\cal{H}}_{+}$ is the usual physical states space and ${\cal{H}}_{-}={\cal{H}_{+}^*}$.}

It is thoroughly accepted that in quantum electrodynamics and in the presence of gauge invariance, the covariant canonical quantization is accessible only through the Gupta-Bleuler formalism. Inspired by the success of this formalism in quantum electrodynamics and regarding the deep analogy with QED,\footnote{The transformation "$\phi \longrightarrow \phi + constant$", viewed as a gauge transformation, is a symmetry for the dS minimally coupled massless scalar field Lagrangian.} recently, a causal and fully covariant construction of the minimally coupled massless scalar quantum field on dS space has been obtained; the so-called Krein-Gupta-Bleuler structure or the "Krein Space Quantization" \cite{de Bievre6230,Gazeau1415,Garidi}. Indeed, it has been shown that to pursue the guideline of the covariance of the full theory under the full de Sitter group $\mbox{SO}_{0}(1,4)$ visualized along the lines proposed by the Wightman-G\"{a}rding in their original work \cite{Wightman129}, utilizing the Krein space construction is indispensable. Actually, an appropriate selection of the space of solutions of the classical field equation assures the causality and the covariance of the theory. Interestingly, in contrast to prior treatments of the problem, this formalism is not contaminated with infrared divergence.

On the other side, as is renowned, the behavior of the graviton two-point function in de Sitter space completely resembles that for the minimally coupled massless scalar field \cite{Ford1601}. This similarity is indeed the origin of the advent of infrared divergences in the graviton two-point function \cite{Allen813,Floratos373,Antoniadis1037}. In this regard, the fully dS covariant quantization of the minimally coupled scalar field through the KGB quantization method may provide a proposal to construct the fully dS covariant and infrared-free graviton two-point function on de Sitter background; at least in the absence of graviton-graviton interaction, the causality and the covariance of the theory are guaranteed. The formalism, therefore, is free of any infrared divergence. For detailed discussions see Ref. \cite{Hamed} (and also \cite{Dehghani064028,Pejhan2052,Rahbardehghan}).

In addition, in this construction interestingly no infinite term appears in the calculation of mean values of the energy-momentum tensor and the vacuum energy of the free field vanishes without any reordering nor regularization. Actually, the covariant renormalization of the energy-momentum tensor is automatically obtained (Wald axioms are well preserved) \cite{Gazeau1415}. The behaviour of Krein quantization in Minkowski space-time, especially when interaction is present, has been also studied in Ref. \cite{Garidi}. In fact, a successful quantization method in de Sitter space-time must lead to the common results when applied in the usual flat case. In this respect, with regard to the condition that preserves the unitarity, it has been shown that the theory is capable of retrieving the results of (Hilbert space) QFT's counterpart in the Minkowskian limit. The only difference between the Krein approach and the usual QFT in flat space-time is restricted to the vanishing of the free field vacuum energy.

As discussed earlier, the inclusion of black hole radiation is mandatory for any consistent quantum field theory in curved space-time. Considering the capabilities of the so-called Krein-Gupta-Bleuler quantization formalism, it may be a good candidate for such a consistent theory, at least in de Sitter space. In this regard, the main purpose of this paper is to describe how thermodynamical aspects of black holes can be perused in the context of the Krein construction, even regarding that the free field vacuum expectation value of the energy-momentum tensor is zero. The layout of the paper is as follows. In section II, the Krein-Gupta-Bleuler structure lying behind the minimally coupled massless scalar field is described. By using this construction the symmetry breaking is prevented altogether. Section III is devoted to studying non-rotating black holes thermodynamics through the KGB quantization. It is explicitly shown that the obtained result is in complete agreement with other works. Finally, a brief conclusion and an outlook for further investigations have been presented.

\section{The KGB structure behind de Sitter minimally coupled massless scalar field}
As already mentioned, respecting the minimal physical requirements for an acceptable quantization scheme, the usual Hilbertian canonical quantization fails in many situations. For instance, in quantum electrodynamics and in the presence of gauge invariance, the covariant canonical quantization is accessible only through the Gupta-Bleuler formalism. How Gupta-Bleuler method is utilized in QED demands clarification. In this regard, consider the space $V_g$ as the space of gauge states (longitudinal photon states), $V$ as the space of positive frequency solutions of the field equation on which the Lorentz condition is satisfied and $V'$ as the space of all positive frequency solutions, including un-physical states, $V_g\subset V\subset V'$. The Fock space is built on $V'$, which is not a Hilbert space, but an indefinite inner product space. It is Poincar\'{e} and locally and conformally invariant. Indeed, all of these three spaces bear representations of the Poincar\'{e} group, however, $V_g$ and $V$ are not invariantly complemented. The quotient states space, $V/V_g$, is the physical one-photon states space up to a gauge transformation (for more mathematical details, one can refer to \cite{BinegarGUPTA,GazeauGUPTA}).

Interestingly, the same scheme will appear in the case of the dS massless minimally coupled scalar field; its Lagrangian
$${\cal{L}}=\sqrt{|g|}\partial_\mu \phi \partial^\mu \phi,$$
is invariant under the global transformation, $\phi\rightarrow\phi + constant$, which behaves in the same manner as a gauge transformation. As a result, an identical fulfillment of the generalized Gupta-Bleuler construction for dS minimally coupled massless scalar field would not be surprising; the role of $V_g$ is played by the set $\cal{N}$ of constant functions. While $\cal{K}$, the physical one-particle space, is defined as a space of positive frequency solutions of the field equation. In this space,  the Klein-Gordon inner product is degenerate though positive. As is well-known, the canonical quantization constructed by a degenerate space of solutions unavoidably breaks the covariance of the field. Consequently, a larger space $K$, a non-degenerate invariant space of solutions which accepts $\cal{K}$ as an invariant sub-space, must be constructed. The so-called Gupta-Bleuler triplet is constituted of these spaces together with ${\cal{N}}$ as: ${\cal{N}}\subset{\cal{K}}\subset K$. It is proved that $K$, called the total space, is a Krein space; the direct sum of ${\cal{H}}_+$ and ${\cal{H}}_-$ \cite{de Bievre6230,Gazeau1415}.

As always in a Gupta-Bleuler quantization formalism, the quantum field is strictly defined as an operator-valued distribution on a Fock space constructed upon the total space, where although the Klein-Gordon inner product is non-degenerate, it is not positively definite. Note that, the total space $K$ is not the space of physical states, because the negative norm states are existing therein. As a result, selecting the physical states sub-space is needed to guarantee a reasonable interpretation of the theory. This selection for de Sitter space-time can be performed by requesting that the physical states be positive frequencies with respect to the conformal time.\footnote{Note that, in Minkowski space-time, we are able to pick out the positive and negative frequency modes because of the existence of time-like Killing vector associated with the Poincar\'{e} symmetry. Generally, in curved space-time, Poincar\'{e} symmetry is lost and there is no preferred set of modes. However, in de Sitter space, for which there is no time-like Killing vector, the physical states can be characterized by demanding that be positive frequencies regarding the conformal time.} It turns out that the set of physical states is characterized \emph{stricto sensu} by a Hilbert space bearing the unitary irreducible representations of the de Sitter group, called the quotient space ${\cal{K}}/{\cal{N}}$. It is shown that this formalism provides a causal, fully covariant, and infrared-free dS massless minimally coupled scalar quantum field \cite{de Bievre6230,Gazeau1415}.

We must emphasize that, due to the fact that the field is built on a non-Hilbertian Fock space, there is no contradiction with the result of Allen \cite{Allen3136,Allen3771}. Actually, in the context of the Krein formalism, the field acts on a space of states which has the Fock space structure and comprises both positive and negative norm states
\begin{equation} \label{FE1} \varphi(x)=\frac{1}{\sqrt{2}}[\varphi_{+}(x)+\varphi_{-}(x)],\end{equation}
where
\begin{eqnarray}\varphi_{+}(x)&=&\sum_{k\geq0}(a_k\phi_k(x) + a^\dag_k\phi^\ast_k(x)),\nn\\
\varphi_{-}(x)&=&\sum_{k\geq0}(b^\dag_k\phi_k(x) + b_k\phi^\ast_k(x)).\end{eqnarray}
Note that, $\Box_H \phi(x) = 0$ and $\Box_H = g_{\mu\nu}^{dS} \nabla^\mu \nabla^\nu$ is the Laplace-Beltrami operator in dS space (for mathematical details see \cite{Gazeau1415}). The positive mode $\varphi_{+}(x)$ is the same scalar field that Allen has used \cite{Allen3136,Allen3771}. This canonical quantization formalism differs with the standard quantum field theory only on the requirement of the following commutation relations
\begin{equation}\label{ccr}
[a_k,a^\dagger_{k'}]=\delta_{kk'},\;\;\; [b_k,b^\dagger_{k'}]=-\delta_{kk'},
\end{equation}
and the other commutation relations vanish.

The (Krein-)Fock vacuum, $|0\rangle$, is determined as follows
\begin{equation}\label{Fockvacuum}
a_k|0\rangle=0,\;\;\; b_k|0\rangle=0.
\end{equation}
The vacuum is invariant under the action of de Sitter group \cite{de Bievre6230,Gazeau1415}. More exactly, the Fock vacuum is unique (Bogoliubov transformations do not alter it) and normalizable. This is not, however, worrying because in this formalism not only the vacuum but the field itself is different. To make this clear, here, we describe how observables are determined in the Gupta-Bleuler type structures. The observables are defined so that possessing this property that they do not "see" the gauge states. The field is indeed gauge dependent, thus it is not an observable. With respect to the fact that $\mu$ stands for global coordinates, the procedure of constructing the physically interesting observables on the total space such as the energy-momentum tensor, therefore, is performed by using operators $\partial_\mu$ \cite{Gazeau1415}. While, as always in a Gupta-Bleuler type formalism, the mean values of observables would be evaluated only with physical states
\begin{equation}
|\vec{k}\rangle = |{k}_1^{n_1}...{k}_j^{n_j}\rangle = \frac{1}{\sqrt{n_1!...n_j!}}(a_{k_1}^\dag)^{n_1}...(a_{k_j}^\dag)^{n_j}|0\rangle,
\end{equation}
which are in fact the elements of ${\cal{K}}/{\cal{N}}$. We should emphasize the fact that the invariance of the Fock vacuum under Bogoliubov transformations does not imply that these transformations are no longer valid in this construction. More precisely, any physical state depends on the observer and also on the chosen space-time; in view of an accelerated observer in Minkowski space, the physical states are different from those in view of an inertial observer (Unruh effect), while for both cases, the same representation of the field is utilized (it is independent of Bogoliubov transformations which are merely a simple change of the space of the physical states) \cite{Garidi}. Indeed in the KGB context, instead of multiplicity of vacua, there are several possibilities for the physical states space, so the usual ambiguity about vacua is not suppressed but displaced.

Here, to see the points, the energy-momentum tensor operator $T_{\mu\nu}$, as an operator on the full space of states (it is not positive definite), is considered. The starting point to evaluate the mean values of the energy-momentum tensor on the sub-space of physical states, $\langle \vec{k}|T_{\mu\nu}| \vec{k}\rangle$, is
\begin{eqnarray}
\langle \vec{k}|\partial_\mu\varphi(x)\partial_\nu\varphi(x)|\vec{k}\rangle = \sum_k \partial_\mu\phi_k(x)\partial_\nu\phi_k^\ast(x)\hspace{1.5cm}\nn\\
- \partial_\mu\phi_k^\ast(x)\partial_\nu\phi_k(x) + 2 \sum_i n_i \Re \Big(\partial_\mu\phi_{k_i}^\ast(x)\partial_\nu\phi_{k_i}(x)\Big). \end{eqnarray}
Note that, the second term with the minus sign comes from the terms of the field containing $b_k$ and $b_k^\dag$. Interestingly, due to the unusual presence of this term, there is an automatic renormalization of the $T_{\mu\nu}$'s (no infinite term appears), and we have
\begin{equation} \langle \vec{k}|\partial_\mu\varphi(x)\partial_\mu\varphi(x)|\vec{k}\rangle = 2 \sum_i n_i \partial_\mu\phi_{k_i}^\ast(x)\partial_\mu\phi_{k_i}(x). \end{equation}
Therefore, the positivity of the energy for any physical state $|\vec{k}\rangle$, which assures a reasonable physical interpretation of the formalism, is guaranteed; $\langle \vec{k}|T_{00}| \vec{k}\rangle\geq0$ ($ = 0 \Leftrightarrow |\vec{k}\rangle=|0\rangle$).

This automatic renormalization of the $T_{\mu\nu}$'s interestingly fulfills the so-called Wald axioms \cite{Gazeau1415}:
\begin{itemize}
\item{The causality and covariance are guaranteed since the constructed field is causal and covariant.}
\item{The above calculation reveals that the usual results for physical states are preserved.}
\item{The foundation of the above calculation is as follows
$$[b_k,b_k^\dag]=-1,$$
accordingly, we have
$$a_ka_k^\dag+a_k^\dag a_k+b_kb_k^\dag+b_k^\dag b_k=2a_k^\dag a_k+2b_k^\dag b_k.$$
So, by applying the method to physical states (on which $b_k$ vanishes), one can easily see that the procedure is equivalent to reordering.}
\end{itemize}

Indeed, the method provides an automatic and covariant renormalization of the energy-momentum tensor, for which, all components of the $<T_{\mu\nu}>$ vanish in the Krein-Fock vacuum. It is actually free of any conformal anomaly in the trace of the energy-momentum tensor. In this regard, it seems that we face with a very different renormalization scheme from the other ones which all exhibit this anomaly. More precisely, in the standard quantum field theory, where the Fock space is constructed on a Hilbert space, any instruction for renormalizing the energy-momentum tensor, which is consistent with Wald axioms, must produce exactly the trace, modulo the trace of a conserved local curvature term \cite{Wald1477}. While, in the KGB context, covariance and conformally covariance are indeed preserved in a rather strong sense, thus there is no surprise that the trace anomaly vanishes (it can appear only through the conformal anomaly). Again, there exists no non-trivial covariant positive type construction for the de sitter minimally coupled quantum scalar field, and hence, this result is nothing but another formulation of Allen's theorem cited above. Moreover, we must express that though the trace anomaly is not present in the context of the KGB structure, but Wald axioms are well preserved.

\emph{Remarks on the renormalization:} In the semi-classical approach to quantum gravity we are involved with calculating the expectation value of $T_{\mu\nu}$ in a special vacuum \cite{Birrell1982,Wald1984}. In this frame, however, the standard formalism of renormalization comprises intricacy and somewhat ambiguity. For instance, no conceptual support exists for a local measure of energy-momentum of some given state without any reference to any global structure. Moreover, energy is the source of curvature, and so contrary to non-gravitational physics, any subtraction of unwanted part of energy (even though it is infinite) is not allowed. Therefore, it seems that a more elaborate renormalization scheme is needed. In this regard, the features of the KGB structure altogether may be helpful for further investigations of QFT in curved space-time.

In the next section, respecting the intrinsic specifications of the Krein quantization method, we examine the method by studying black holes thermodynamics, and in particular, calculating Hawking radiation.

\section{Black hole radiation in the context of KGB quantization}
As already mentioned, any consistent quantum field theory in curved space-time must include black hole radiation. Therefore, the study of black holes thermodynamics through the Krein-Gupta-Bleuler construction, as a respectful quantization scheme in de Sitter space-time, is of great importance. In this regard, by evaluating the features and capabilities of the Krein formalism in one side and considering black holes intrinsic specifications on the other side, we have provided the procedure for investigating thermodynamical aspects of black holes in the context of the KGB structure, which is to be described in this section.

In a previous work \cite{SH1408}, we studied a basic but significant case; conformally coupled massless scalar field on dS background subjected to Dirichlet boundary condition on a curved brane. It has been shown that, due to the existence of the boundary condition and by studying Casimir energy-momentum tensor induced by it, the aspects of the gravitational thermodynamics have been emerged in the spirit of the Krein-Gupta-Bleuler approach, even possessing this particular property that $<0|T_{\mu\nu}|0>$ of the free theory is zero. This basic idea (employing various boundary conditions and evaluating related Casimir stress tensors) can be straightforwardly applied to more sophisticated cases like black holes; it is renowned that, intrinsically, the gravitational field of black holes produces an effective potential barrier that acts as a good conductor (it conducts well at low frequencies, but as the frequencies increase, its conductivity diminishes) \cite{Price2419,Fabbri933}. In this section, pursuing this scheme, we consider a non-rotating black hole and evaluate Casimir energy-momentum tensor induced by the potential barrier for a conformally coupled massless scalar field in the vicinity of the black hole event horizon. For our case, a non-rotating black hole, the peak of the potential barrier is so well localized near $r = 3M$ ($M$ refers to the hole's mass; $c = G = 1$) \cite{Price2419}, so that, the black hole is almost unable to absorb electromagnetic waves when their frequencies are less than $\omega_c\cong  (2/27)^{1/2} M^{-1}$ \cite{Fabbri933}. Indeed, $\omega_c$ is the cut-off frequency for the absorption of electromagnetic and scalar fields by a Schwarzschild black hole.

In the generic black hole background, the study of boundary-induced quantum effects is technically complicated. However, the exact analytical result can be evaluated in the vicinity of the black hole event horizon. In this limit, the black hole geometry may be well approximated by the Rindler-like manifold (for some investigations of quantum effects on the background of Rindler-like space-times see \cite{Bytsenko,Zerbini,Cognola}). In this framework, moreover, due to the gravitational red-shift, the gravitational field potential barrier may be well approximated by a good conductor with proper acceleration $b^{-1}\cong (3\sqrt{3}M)^{-1}$.\footnote{The general formula for proper acceleration of a particle which is at rest in the gravitational field of a non-rotating black hole is as follows \cite{Brown}: $(1-2M/r)^{-1/2}M/r^2$, while a stationary distant observer will measure $ M/r^2 $. So, the potential barrier localized in the vicinity of $r=3M$ has a non-zero proper acceleration $(3\sqrt{3}M)^{-1}$.} Evaluating Casimir energy-momentum tensor for a conformally coupled massless scalar field in the vicinity of the black hole event horizon, therefore, would be met by applying the results of the accelerated-mirror; the potential barrier is substituted by a plane conductors with proper acceleration $b^{-1}$.

In this regard, a massless scalar field subjected to Dirichlet boundary condition in Minkowski space-time induced by an infinite plane conductor (uniformly accelerated normal to itself) is considered. At the beginning, let us illustrate the Krein quantum field operator for the massless scalar field in Minkowski space-time satisfying the following field equation
\begin{equation}\Box \phi(x)=0, \end{equation}
where $\Box \equiv g^{\mu \nu} \partial_\mu \partial_\nu$ and $g_{\mu \nu} = (-\;+\;+\;+)$. The inner product of a pair of its solutions is defined by
\begin{equation}(\phi_1,\phi_2)=i\int_t (\phi_1(x) \, {\mathop{\partial_t}\limits^\leftrightarrow } \,\phi^*_2(x)) d^3 x.\end{equation}
As explained, the field operator $\varphi$ in Krein quantization is $\varphi = \frac{1}{\sqrt{2}}({\varphi}_+ + {\varphi}_-)$, in which
\begin{eqnarray}\label{fk}
\varphi_+(x)&=&\int d^3 \vec{k}\;[a(\vec{k})\phi(\vec{k},x)+a^\dag(\vec{k})\phi^\ast(\vec{k},x)],\nn\\
\varphi_-(x)&=&\int d^3 \vec{k}\;[b(\vec{k})\phi^\ast(\vec{k},x)+b^\dag(\vec{k})\phi(\vec{k},x)],
\end{eqnarray}
$\phi (\vec{k},x)=({4\pi\omega})^{-1/2}{e^{i\omega t - i\vec{k}\cdot\vec{x}}}$ and $[a(\vec{k}), a^\dagger(\vec{k}')] = \delta (\vec{k}-\vec{k}'),\;\; [b(\vec{k}), b^\dagger(\vec{k}')] = -\delta (\vec{k}-\vec{k}')$, the other commutation relations are zero. The Fock vacuum state $|0\rangle$ is defined by $a(\vec{k}) |0\rangle=0,\; b(\vec{k}) |0\rangle=0$.

In the KGB context, due to the presence of un-physical states (the states with negative energy), when interacting field theories are taken into account, the unitarity of the S-matrix must be preserved. This would be guaranteed by following the so-called unitarity condition; to see the points, consider the following Lagrangian density with a potential term $V$
\begin{equation}{\cal{L}}=g^{\mu\nu}\partial_\mu\phi\partial_\nu\phi - V(\phi).\end{equation}
Here, $K={\cal{H}}_{+}\oplus {\cal{H}}_{-}$ characterizes the free field Krein-Fock space, while in the absence of gauge invariance, the space of physical states ${\cal{H}}_{+}$ is closed and positive. Let $\Pi_+$ be the projection over ${\cal{H}}_+$
$$\Pi_+ = \sum_{\{\alpha_+\}} |\alpha_+><\alpha_+|,\;\;\;\;\; |\alpha_+>\;\in {\cal{H}}_+.$$
So, considering the field operator, one has
\begin{equation}\label{unicon} \Pi_+ \varphi \Pi_+ |\gamma > = \left\{\begin{array}{rl} &\varphi_+ |\gamma >, \;\;\;\;\;\; \mbox{if}\; |\gamma>\;\in {\cal{H}}_+ \vspace{2mm}\\\vspace{2mm} &0,\;\;\;\;\;\;\;\;\;\;\;\;\;\;\;\;\mbox{if}\; |\gamma>\;\in {\cal{H}}_{-} \\\end{array}\right. \end{equation}
Accordingly, instead of a standard choice for the Lagrangian potential term, $V(\varphi)$, we start with $V'(\varphi)\equiv V(\Pi_+\varphi\Pi_+)$ which is the restriction of $V$ to the positive energy modes,
\begin{equation}{\cal{L}}=g^{\mu\nu}\partial_\mu\phi\partial_\nu\phi - V'(\phi).\end{equation}
It is proved that the unitarity of the S-matrix would be achieved through this condition \cite{Garidi}. In this respect, the vacuum effects in the theory only involve the interacting vacuum. Note that, the so-called radiative corrections, in the Minkowskian limit, are the same in both theories. Indeed, in this limit, vanishing of the free field vacuum energy in the Krein method is the only difference between the Krein approach and the usual one.

So, by imposing physical boundary condition on the field operator, only physical states are affected. The field operator then would be
\begin{eqnarray}\label{effected}
\varphi(x)=\sum_d [a({\vec{k}}_d) \phi(\vec{k}_d,x)+a^\dag({\vec{k}}_d)\phi^\ast(\vec{k}_d,x)]\nn\\ + \int d^3 \vec{k}\; [b(\vec{k})\phi^\ast (\vec{k},x)+b^\dag (\vec{k})\phi (\vec{k},x)],
\end{eqnarray}
here $\vec{k}_d$ are the eigen-frequencies of the system under consideration.

As mentioned above from now on, we exploit the accelerated (Rindler) coordinates defined by the coordinate transformation: $t=\sigma\sinh\eta$, $x=\sigma\cosh\eta$, which cover the region $|x|>|t|$ of Minkowski space. The line element is as $ds^2 = -\sigma^2d\eta^2+d\sigma^2+dx^2$, with $x=(y,z)$. The curves $\sigma=\mbox{constant}$, $x=\mbox{constant}$ are worldlines of constant proper acceleration $\sigma^{-1}$ and the surface $\sigma=b$ represents the trajectory of the barrier which has a proper acceleration $b^{-1}$.

The starting point to calculate the vacuum expectation value of the energy-momentum tensor in view of the accelerated observer in Minkowski space-time is \cite{Candelas}
\begin{eqnarray} \label{st}
<0|T_\mu^\nu|0> =-i\lim_{x'\rightarrow x} (\frac{2}{3}\nabla_{\mu}\nabla^{\nu'}- \frac{1}{3}{\nabla_{\mu}}\nabla^{\nu}\hspace{1cm}\nn\\
- \frac{1}{6}g_\mu^\nu\nabla_{\alpha}\nabla^{\alpha'})G(x,x'),
\end{eqnarray}
where $G(x,x')$ is the Feynman Green function that with respect to the quantum field of the theory can be decomposed into two parts, physical and un-physical parts (the point is $(\phi_k,\phi_{k'}^*)= 0$ and $[a(k), b^\dagger(k')] = [a^\dagger(k),b^\dagger(k')] = 0$). Accordingly, we have
$$ G(x,x')= G_{+}(x,x') + G_{-}(x,x').$$

As mentioned before, with regard to the unitarity condition, only the physical part of the theory, $G_{+}(x,x')$, would be subjected to Dirichlet boundary condition. The corresponding propagator, the positive frequency Feynman Green function subjected to Dirichlet boundary condition, has been  computed in Ref. \cite{Candelas}
\begin{eqnarray}\label{DP}
G_{+}(x,x')= G_{0}(x,x') \hspace{5cm} \nn\\
- \frac{i}{\pi}\int \frac{d\nu}{2\pi} \exp [-i\nu(\eta-\eta')]\int \frac{d^2k}{(2\pi)^2}\exp [ik\cdot (x-x')] \nn\\
\times\frac{K_{i\nu}(e^{i\pi}kb)}{K_{i\nu}(kb)} K_{i\nu}(k\sigma) K_{i\nu}(k\sigma'),\;\;
\end{eqnarray}
in which $K_{i\nu}(k\sigma)$ is the modified Bessel function of imaginary order and
\begin{eqnarray}
G_{0}(x,x') = \frac{i}{\pi}\int \frac{d\nu}{2\pi} \exp [-i\nu(\eta-\eta')]\hspace{2.5cm}\nn\\
\times \int \frac{d^2k}{(2\pi)^2}\exp [ik\cdot (x-x')] K_{i\nu}(k\sigma_>) K_{i\nu}(e^{i\pi}k\sigma_<).\hspace{0.2cm}
\end{eqnarray}
Note that, all divergences are contained in $G_{0}$, which indeed corresponds to the positive frequency Feynman propagator for a massless scalar free field on the entire Minkowski manifold.

On the other side, the un-physical part of the Green function remains unaffected, and accordingly, would be written in the sense of $G_{0}$ in the following form \cite{takookIJTP}
\begin{eqnarray}\label{DP'} G_{-}(x,x') =  \Big(G_{0}(x,x')\Big)^{\ast}. \end{eqnarray}

Now, by substituting (\ref{DP}) and (\ref{DP'}) into (\ref{st}) and after a straightforward computation, one can obtain
\begin{equation}\label{diagT} <T_\mu^\nu>=\mbox{diag}(A,B,C,C), \end{equation}
in which
$$ A(\sigma)=\frac{1}{12\pi^3}\int_0^{\infty} d\nu\int_0^{\infty} dk\;k\frac{K_{i\nu}(e^{i\pi}kb)}{K_{i\nu}(kb)}\Big[k^2K_{i\nu}^{'2}(k\sigma)$$
$$ +\frac{2k}{\sigma}K_{i\nu}(k\sigma)K'_{i\nu}(k\sigma)+\left( \frac{5\nu^2}{\sigma^2}+k^2 \right)K_{i\nu}^2(k\sigma) \Big],$$

$$ B(\sigma)=-\frac{1}{12\pi^3}\int_0^{\infty} d\nu\int_0^{\infty} dk\;k\frac{K_{i\nu}(e^{i\pi}kb)}{K_{i\nu}(kb)}\Big[ 3k^2K_{i\nu}^{'2}(k\sigma)$$
$$ +\frac{2k}{\sigma}K_{i\nu}(k\sigma)K'_{i\nu}(k\sigma)+3\left( \frac{\nu^2}{\sigma^2}-k^2 \right)K_{i\nu}^2(k\sigma) \Big],$$

$$ C(\sigma)=-\frac{1}{12\pi^3}\int_0^{\infty} d\nu\int_0^{\infty} dk\;k\frac{K_{i\nu}(e^{i\pi}kb)}{K_{i\nu}(kb)}\Big[ -k^2K_{i\nu}^{'2}(k\sigma)$$
$$ +\left( \frac{\nu^2}{\sigma^2}+2k^2 \right)K_{i\nu}^2(k\sigma) \Big].$$
It is trace-free, $A+B+2C=0$, and conserved, $A=\frac{d}{d\sigma}(\sigma B)$. It is worth mentioning that, in Ref. \cite{Candelas} by P. Candelas \textit{et al}., the above regularized form was calculated by subtracting the value that it would have if evaluated relative to the Minkowski vacuum ($G_{0}(x,x')$). In the Krein construction, however, the vacuum expectation value of $T_\mu^\nu$ is automatically regularized.

Now, in analogy with the suggested process by P. Candelas \textit{et al}., one can evaluate the above integrals. In this respect, far from the barrier, in the frame of an observer with proper acceleration $\sigma^{-1}$ ($\sigma/b \rightarrow\infty$), the asymptotic form of $<T_\mu^\nu>$, which is of particular interest in our model to simulate schwarzschild black holes, would be
\begin{equation}\label{MM} <0|T_\mu^\nu|0>\sim \frac{-1}{2\pi^2\sigma^4} \int_0^\infty \frac{\nu^3 d\nu}{e^{2\pi\nu}-1}diag(-1,\frac{1}{3},\frac{1}{3},\frac{1}{3}). \end{equation}
It presents a negative energy density with a Planckian spectrum with the temperature $T=(2\pi\sigma)^{-1}$, corresponding to the absence from the vacuum of black-body radiation. This asymptotic form is independent of the acceleration of the barrier, and depends only on the local acceleration.

Therefore, respecting the gravitational analogy, for an observer who is at rest next to the horizon ($r\approx r_0$) with proper acceleration $(M/r_0^2)(1-2M/r_0)^{-1/2}$, at adequately large distances to the potential barrier, the vacuum expectation value of the energy-momentum tensor depends purely on the local gravitational field, and accordingly, in this frame work, the observer will discover a negative flux of black-body radiation with the temperature
\begin{equation}\label{nf}T_0=(1/2\pi)(M/r_0^2)(1-2M/r_0)^{-1/2}.\end{equation}
[Before coming back to this point, it is worth mentioning that due to the initial flow of negative Casimir energy \cite{Nugayev}, the mass of the hole diminishes. Thus, the acceleration $b^{-1}$ of the potential barrier is not uniform. However, it is not a difficulty to apply the accelerated-mirror results since
$$\sigma/b = \frac{r_0^2(1-2M/r_0)^{1/2}}{3\sqrt{3}M^2} \rightarrow\infty,\;\;\;\mbox{if} \;\; M\rightarrow 0.$$
This is why at late time we can use (\ref{MM}).]

In the perspective of an observer far from the black hole - at future infinite - when the particle is infinitesimally close to the event horizon, this negative flow (\ref{nf}) through the horizon shows up as a positive flux of particles at infinity with
\begin{equation}T=(1/2\pi)(M/r_0^2),\end{equation}
$M/r_0^2$ is the magnitude of the acceleration measured by an observer at infinity. Interestingly, considering that $k$ is the surface gravity on the horizon of a schwarzschild black hole, $k=\frac{1}{4M}$, then we have
$$T=k/2\pi.$$
This result is directly analogous to the black hole thermal radiation computed by Hawking \cite{Hawking199}.

\section{Conclusion and outlook}
This paper is part of a series of papers with the aim of developing a complete self-consistent formalism for the treatment of quantum field theory in curved space-time. In this path, the study of thermodynamical properties of black holes in the context of any respectful quantization scheme in curved space, is a crucial step to take. As already pointed out, an intrinsic property of black holes, the effective potential barrier created by the gravitational field, has been exploited to perform the inspections. One of the main strengths of this view is the consistency of the procedure it provides of the black hole radiation with the Krein specifications. Following these lines, by obtaining the exact usual result for the temperature of black hole radiation, we have indeed shown that the theory is capable of being used for further investigations of thermodynamical aspects of black holes.

The procedure we have utilized to achieve this goal, investigating boundary conditions and physical achievements they bring about, has also a great significance in brane-world scenarios, which are of considerable interest in both particle physics and cosmology \cite{Randall3370,Wasserman1682,Elizalde063515}. The presence of boundaries and propagating fields in the bulk is indeed an intrinsic feature of these scenarios. These models, therefore, will give Casimir-type contributions to the vacuum expectation values of physical observables, which have to be taken into account in the self-consistent formulation (for detailed reviews of the Casimir effect refer to \cite{Elizalde}). We hope that this study to give a description of the black hole radiation mechanism through the Krein-Gupta-Bleuler construction, along with the capabilities of the method stated before, can prepare the stage for further developments, particularly extension of the theory into the relevant subjects of dS/CFT correspondence such as quantum cosmology and brane-world inflationary.

\section*{Acknowledgments}
The authors would like to thank professor J. P. Gazeau for instrumental discussions which have been quite effective in development of the approach.

\end{document}